\documentclass[%
    aps,
    prd,
    reprint,
    letterpaper,
    superscriptaddress,
    preprintnumbers,
    floatfix,
    twocolumn,
    nofootinbib,
    hyphens
]{revtex4-2}  
\usepackage{microtype}
\usepackage{graphicx} 
\usepackage{comment}    
\usepackage{dcolumn}   
\usepackage{bm}       
\usepackage{amssymb}  
\usepackage{subfigure}
\usepackage{multirow}
\usepackage{slashed}
\usepackage{placeins}
\usepackage{epstopdf}
\usepackage{mathtools}
\usepackage{color}
\usepackage{tikz,pgfplots}
\pgfplotsset{compat=newest}
\usepackage{enumerate,comment}
\usepackage{setspace}
\usepackage{amsmath}
\usepackage{bm}               
\usepackage{amssymb}
\usepackage{amsfonts}
\usepackage{mathtools}
\usepackage{dsfont}
\usepackage{commath}
\usepackage{physics}
\usepackage[normalem]{ulem}
\usepackage{tensor}
\usepackage{amsfonts}    
\usepackage{amsmath}
\usepackage{dcolumn}\usepackage{tabularx}
\usepackage{listings}
\usepackage{adjustbox}
\usepackage{tikz}
\usetikzlibrary{quantikz}
\makeatletter
\DeclareRobustCommand{\rvdots}{%
  \vbox{
    \baselineskip4\p@\lineskiplimit\z@
    \kern-\p@
    \hbox{.}\hbox{.}\hbox{.}
  }}
\makeatother
\usepackage{amsmath}
\pdfmapfile{+sansmathaccent.map}
\renewcommand{\op}[1]{\hat{#1}}

\renewcommand{\vec}[1]{\boldsymbol{#1}}
\newcommand{\mom}[1]{\vec{\mathrm{#1}}}

\newcommand{\argmin}{\mathrm{argmin}}
\newcommand{\argmax}{\mathrm{argmax}}
\usepackage{hyperref}  
\hypersetup{breaklinks=true}
\usepackage{algorithm}
\usepackage[compatible]{algpseudocode}
\renewcommand{\algorithmiccomment}[1]{\bgroup\hfill//~#1\egroup}

\definecolor{c1}{RGB}{206,0,0}
\definecolor{c2}{RGB}{249,149,0}
\definecolor{c3}{RGB}{153,0,210}
\definecolor{c4}{RGB}{0,109,219}
\definecolor{c5}{RGB}{0,146,146}
\definecolor{c6}{RGB}{255,109,182}
\pgfplotscreateplotcyclelist{cycle1}{
{c1}, {c2}, {c3}, {c4}, {c5}, {c6}
}
\begin{document}
\preprint{
\vbox{\hbox{MIT-CTP/{5459}}}}
\preprint{
\vbox{\hbox{ADP-22-26/T1197}}}
\title{Strategies for quantum-optimized construction of interpolating operators in\\ classical simulations of lattice quantum field theories}
\author{A.~Avkhadiev}\affiliation{Center for Theoretical Physics, Massachusetts Institute of Technology, Cambridge, MA 02139, U.S.A.}
\affiliation{DOE Co-Design Center for Quantum Advantage}
\author{P.~E.~Shanahan}
\affiliation{Center for Theoretical Physics, Massachusetts Institute of Technology, Cambridge, MA 02139, U.S.A.}
\affiliation{DOE Co-Design Center for Quantum Advantage}
\affiliation{NSF AI Institute for Artificial Intelligence and Fundamental Interactions}
\author{R.~D.~Young}\affiliation{CSSM, Department of Physics,
  University of Adelaide, Adelaide SA 5005, Australia}
\begin{abstract}
    It has recently been argued that noisy intermediate-scale quantum computers may be used to optimize interpolating operator constructions for lattice quantum field theory (LQFT) calculations on classical computers. 
    Here, two concrete realizations of the method are developed and implemented. 
    The first approach is to maximize the overlap, or fidelity, of the state created by an interpolating operator acting on the vacuum state to the target eigenstate. 
    The second is to instead minimize the energy expectation value of the interpolated state.
    These approaches are implemented in a proof-of-concept calculation in (1+1)-dimensions for a single-flavor massive Schwinger model
    to obtain quantum-optimized interpolating operator constructions for a vector meson state in the theory. 
    Although fidelity maximization is preferable in the absence of noise due to quantum gate errors, it is found that energy minimization is more robust to these effects in the proof-of-concept calculation. 
    This work serves as a concrete demonstration of how quantum computers in the intermediate term might be used to accelerate classical LQFT calculations.
\end{abstract}
\maketitle
\section{
    Introduction 
    \label{sec:intro}
}
Lattice quantum field theory (LQFT) is a non-perturbative framework to study quantum field theories in a systematically improvable way \cite{Wilson:1974sk, Osterwalder:1977pc}.
In particular for study of the theory of strong nuclear interactions, quantum chromodynamics, LQFT has served as an indispensable tool \cite{Kogut:1982ds, Creutz:2011ati, Lin:2017snn, Detmold:2019ghl, Bazavov:2019lgz, USQCD:2019hyg, Kronfeld:2019nfb, Cirigliano:2019jig, USQCD:2019hee, ParticleDataGroup:2020ssz, USQCD:2022mmc, Amoroso:2022eow}. However, certain aspects of quantum systems, such as real-time dynamics or observables plagued by the sign problem, are difficult or impossible to address with LQFTs formulated in Euclidean space-time and realized with classical computing; it has long been argued that quantum computing may provide a path towards such calculations~\cite{Manin, Feynman1982SimulatingComputers, doi:10.1126/science.273.5278.1073, Wiesner:1996xg, Zalka:1996st, Meyer:1996kt, Boghosian:1996qd, Abrams:1997gk, Preskill:1997dt, Terhal:1998yh, Abrams:1998pd, Ortiz:2000gc, PhysRevA.65.042323, lens.org/055-737-950-838-184}.
This hope has inspired sustained efforts and developments: in experimental design proposals and realizations of quantum simulators, in the development of algorithms for quantum simulation and their proof-of-concept implementations, and in formulations of LQFTs for quantum calculations~\cite{Steane:1997kb, doi:10.1126/science.1177838, Georgescu:2013oza, osti_1631143, Preskill2018QuantumBeyondb, Preskill:2018fag, Joo2019StatusBeyond, Banuls:2019bmf, PRXQuantum.2.017003, Klco:2021lap, Bauer:2022hpo, Tews:2022yfb}.
Many of these efforts are concentrated on frameworks where the calculations are carried out principally on a quantum computer, with classical computers playing a supporting role; others consider frameworks with auxiliary quantum-computer calculations at select points in a conventional, classical-computer LQFT workflow.
Such hybrid approaches will not shift the paradigm and scope of LQFT in the same way as inherently quantum approaches that may be enabled by large-scale fault-tolerant quantum computation; nevertheless, the hybrid approaches may offer a practical advantage and accelerate classical QFT studies in the intermediate term.

As proposed in Ref.~\cite{Avkhadiev:2019niu}, interpolating operator optimization is one task that could be performed on quantum computers as a path towards accelerating classical LQFT calculations. Interpolating operators in LQFT are constructed to approximately map the vacuum state of a given theory to the ground state with a particular set of quantum numbers, thus ``interpolating'' between the states. A perfect interpolating operator would realize the exact map; in practice, however, interpolated states --- quantum states created by the action of interpolating operators on the vacuum state --- are linear superpositions of all states with the quantum numbers of the operator, and ground states are obtained from interpolated states by Euclidean time-evolution, which exponentially suppresses excited-state contributions.
Improved interpolating operators can enable more precise extraction of physics quantities for a fixed computational cost, for example by reducing statistical variance in those quantities or by reducing excited-state contamination and enabling ground-state dominance at earlier Euclidean times.
In this work, two concrete strategies are formulated for interpolating operator optimization on quantum devices.
The first strategy is to maximize the fidelity of the interpolated state to the ground state. 
The second strategy is to minimize the energy expectation value of the interpolated state. 
Implementations of both strategies are demonstrated in 1+1 dimensions for the massive single-flavor Schwinger model using a quantum-computer simulator. 
It is shown that, while maximizing fidelity is preferable in principle, its implementation, which requires deeper quantum circuits with a larger number of entangling gates and twice as many qubits, makes the strategy particularly susceptible to gate errors.
In comparison, while minimizing energy expectation value is less general and less optimal in principle, its implementation is more robust to gate errors.

The rest of this manuscript is organized as follows:
Sec.~\ref{sec:spectroscopy} defines the task of interpolating operator optimization;
Sec.~\ref{sec:strategies} introduces the two strategies for interpolating operator optimization and describes their implementation; Sec.~\ref{sec:schwinger} presents proof-of-concept realizations and results for both strategies;
Sec.~\ref{sec:hybrid-workflow} discusses how this work advances the use of quantum computing within LQFT.
Finally, Section~\ref{sec:conclusion} provides an outlook. 
\section{
    Lattice spectroscopy and interpolating operator optimization 
    \label{sec:spectroscopy}
}
Physical observables in LQFT can be computed from the Euclidean time dependence of correlation functions \cite{Parisi:1983ae}.
For example, the energy of a ground state with a given set of quantum numbers can be computed from a momentum-projected two-point correlation function $C(\mom{p},t)$ of time-local interpolating operators $\op{\mathcal{O}}^\dagger(\mom{p},0)$ and $\op{\mathcal{O}}(\mom{p},t)$ with the quantum numbers of the target ground state $\ket*{E_0}$:
\begin{equation}
\label{eq:twopt}
\begin{aligned}
C(\mom{p},t) 
	&\equiv \left\langle \Omega \right\rvert
		\hat{\mathcal{O}}(\mom{p},t)
		\hat{\mathcal{O}}^\dagger(\mom{p},0)
       \left\lvert \Omega \right\rangle \\
    &= \lvert Z^0_{\mathcal{O}}\rvert e^{-E_{0}(\mom{p})t}\\
    &\quad\times\left(
        1 + \sum_{n>0} 
            \frac{\lvert Z^n_\mathcal{O}\rvert}
                {\lvert Z^0_\mathcal{O} \rvert}
            e^{-(E_{n}(\mom{p})-E_0(\mom{p}))t}
        \right).
\end{aligned}
\end{equation}
Here, $\ket*{\Omega}$ denotes the vacuum state of the theory and $E_{n}(\mom{p})$ denotes the energy of the $n$th excited state with the same quantum numbers as $\ket*{E_0}$. States are normalized non-relativistically, $\braket*{E_{m}(\mom{q})}{E_{n}(\mom{p})} = \delta_{n, m} \delta_{\mom{p},\mom{q}}$. Overlap factors are defined as
\begin{equation}
\label{eq:ovlap}
\begin{aligned}
	\sqrt{Z^{n}_{\mathcal{O}}(\mom{p})} =\bra{E_n(\mom{p})}\op{\mathcal{O}}^\dagger(\mom{p}, 0)\ket{\vphantom{E_n(\mom{p})}\Omega}.
\end{aligned}
\end{equation}
Ground-state dominance at late Euclidean times is revealed in the effective energy function:
\begin{equation}
\label{eq:effective-energy}
    E_{\mathrm{eff}}(t) 
        = \frac{1}{a}\log\frac{ C(\mom{p},t)}{ C(\mom{p},t+a)} \xrightarrow{t \gg 0} E_0(\mom{p}).
\end{equation}

In practice, correlation functions are estimated stochastically by evaluation on a finite set of gauge field configurations $U_{i}$, $i \in \lbrace 1,\ldots,N_{\mathrm{cfg}}\rbrace$: 
\begin{equation}
\label{eq:twopt-mc}
    \langle C(\mom{p},t)\rangle
        = \frac{1}{N_{\mathrm{cfg}}}
            \sum_{i}
                \langle C(\mom{p},t, U_{i})\rangle_{\mathrm{F}},
\end{equation}
where the configurations are distributed as $e^{-S_\mathrm{eff}(U)}$, $S_\mathrm{eff}(U)$ is the effective action of the theory with fermions integrated out, and $\langle\cdot\rangle_{\mathrm{F}}$ denotes Wick-contraction of fermions into propagators. While excited-state contributions to  $\langle C(\mom{p},t)\rangle$ are exponentially suppressed at late Euclidean times, the signal-to-noise ratio diminishes exponentially in the same limit for many observables of interest \cite{Parisi:1983ae, Lepage:1989hd}. The precision with which ground-state observables, such as the ground-state energy $E_0$ or matrix elements of currents in the ground state, can be determined is thus related to the overlap factors $|Z_\mathcal{O}^n|$. In particular, given correlation functions with comparable statistical variance computed from different interpolating operators, those from operators with smaller overlap factors onto excited states suffer less from excited-state contamination at earlier times where statistical fluctuations are smaller, and will thus yield a more precise determination of ground-state observables at fixed statistics.

In this work, {\it interpolating operator optimization} is defined as the tuning of an interpolating operator construction $\mathcal{O}$ such that excited-state contamination in $C(\mom{p},t)$  is minimized.
By design, this is a component of the LQFT workflow which is robust to errors or systematic uncertainties, including any induced by a process based on quantum computation. 
That is, correlation functions constructed using a sub-optimal interpolating operator construction will simply have additional excited-state contamination, while optimizing the construction can lead to exponential gains in precision at fixed statistics. 

Interpolating operators may be optimized with classical computing via conventional methods such as variational \cite{Fox:1981xz, Luscher:1990ck, Michael:1982gb, Michael:1985ne, Blossier:2009kd, Mahbub:2009nr, Detmold:2014hla} and Prony \cite{Fleming:2004hs, Lin:2007iq, Beane:2009kya, Beane:2009gs} approaches. These methods have found great success, but require stochastically estimating correlation functions in Euclidean time and are thus sensitive to the onset and the rate of exponential signal-to-noise suppression as discussed above. As a result, the computational cost of classical interpolating operator optimization is increased in systems where the signal-to-noise problem is more severe and thus where optimized interpolating operators would be most needed, such as multi-baryon systems \cite{Beane:2009gs, Beane:2011zpa, Beane:2011zpa, Inoue:2011ai, Ishii:2012ssm, NPLQCD:2012mex, Yamazaki:2012hi, Orginos:2015aya, Berkowitz:2015eaa, Wagman:2017tmp, Iritani:2018vfn, Francis:2018qch, Junnarkar:2019equ, Horz:2020zvv, Green:2021qol, Amarasinghe:2021lqa, Lyu:2022tsd} and boosted states~\cite{DellaMorte:2012xc, Bali:2016lva, Wu:2018tvt}. In contrast, optimization of interpolating operators on quantum devices, as proposed in Ref.~\cite{Avkhadiev:2019niu}, is not sensitive to the signal-to-noise problem in Euclidean time (but faces other challenges, as detailed in the rest of this manuscript). It is thus possible that in the era {\it after} hybrid-workflow LQFT applications are enabled by improvements in quantum-computing technology, but {\it before} full-scale quantum computers can be practically deployed, quantum optimization of interpolating operators in LQFT may be a practical approach that is complementary to classical approaches.

While it has been outlined in Ref.~\cite{Avkhadiev:2019niu} how one might use quantum devices for interpolating operator optimization, to examine whether the task is tenable with noisy intermediate-scale quantum (NISQ) \cite{Preskill2018QuantumBeyondb} devices in particular, it is important to investigate concrete optimization strategies.
\section{
    Quantum interpolating operator optimization strategies
    \label{sec:strategies}
}
In this work, two strategies for quantum-simulation based interpolating operator optimization are introduced and implemented. 
In both cases, given a set of interpolating operators $\lbrace \mathcal{O}_i \rbrace$ that can be defined and practically implemented in formulations of a given LQFT for both classical and quantum-computer calculations, the goal is to produce an optimal linear combination of all operators in the set. 
That is, the output of the optimization is a vector of coefficients $\vec{\alpha}$, such that the corresponding optimal operator is 
\begin{equation}
\mathcal{O}(\vec{\alpha}) = \sum_{i}\alpha_i\mathcal{O}_i,
\end{equation}
where the dependence of operators on momentum or lattice coordinates is suppressed. 
\paragraph{Maximum-fidelity (MF) strategy.} The first strategy is to find $\vec{\alpha}$ that maximizes the overlap of the interpolated state --- the normalized state created by acting on the vacuum state $\ket*{\Omega}$ with the operator $\op{\mathcal{O}}(\vec{\alpha})$ --- onto the target ground state $\ket*{E_0}$, i.e.,
\begin{equation}
\label{eq:max-fidelity}
\begin{aligned}
\vec{\alpha}_{\mathrm{MF}}
    &= \argmax_{\vec\alpha} 
        \left\lvert\braket{E_0}{\mathcal{O}(\vec{\alpha})}\right\rvert^2 \\
    &= \argmax_{\vec\alpha}
	\frac
		{\lvert Z^{0}_{\mathcal{O}}(\vec{\alpha})\rvert}
		{N_{\mathcal{O}}(\vec{\alpha})},
\end{aligned}
\end{equation}
where 
\begin{equation}
\label{eq:interpolated-state}
\ket{\mathcal{O}(\vec{\alpha})} 
    = \frac{1}{\sqrt{N_{\mathcal{O}}(\vec{\alpha})}}
    \sum_{j} \alpha_j\hat{\mathcal{O}}_j^\dagger\ket{\Omega},
\end{equation}
and the wavefunction normalization is given by $N_{\mathcal{O}}(\vec{\alpha}) = \sum_{i,j} \alpha_i^*\alpha_j \bra{\Omega}\op{\mathcal{O}}_i\op{\mathcal{O}}^\dagger_j\ket{\Omega}$.
The effect of this optimization is to reduce the coefficients $|Z_\mathcal{O}^n|/|Z_\mathcal{O}^0|$ of exponential excited state contamination in Eq.~\eqref{eq:twopt}.
In implementing the MF strategy on a quantum device, $\ket*{\Omega}$ and $\ket*{E_0}$ may be approximated with quantum states obtained, for example, by a variational quantum eigensolver (VQE) \cite{Peruzzo2014AProcessor, McClean_2016}. Once such variational approximations are found for both states, they may be used to estimate Eq.~\eqref{eq:max-fidelity} using the decomposition of $\op{\mathcal{O}}^\dagger_i$ and $\op{\mathcal{O}}_i\op{\mathcal{O}}^\dagger_j$ into a basis of unitary Pauli-string operators: $\op{P}_k \in \lbrace \hat{I}, \hat{\sigma}_x, \hat{\sigma}_y, \hat{\sigma}_z \rbrace^{\otimes n}$, $k \in \{1, 2, \ldots 4^n\}$, where $n$ is the number of qubits required to encode the operators. In particular, the denominator $N_{\mathcal{O}}(\vec{\alpha})$ may be computed using the decomposition 
$\op{\mathcal{O}}_i \op{\mathcal{O}}^\dagger_j = \sum_{k}c_k^{(ij)}\op{P}_k$ as
\begin{equation}
\label{eq:compute-norm}
    N_{\mathcal{O}}(\vec{\alpha}) 
        = \sum_k \sum_{i,j} \alpha_i^*\alpha_j c^{(ij)}_{k} \bra{\Omega} \op{P}_k \ket{\Omega},
\end{equation}
with estimates for $\bra{\Omega} \op{P}_k \ket{\Omega}$ obtained from quantum circuits using Pauli measurements \cite{9259964}. 
Similarly, $\lvert Z^{0}_{\mathcal{O}}(\vec{\alpha})\rvert$ in the numerator may be written using the decomposition $\op{\mathcal{O}}^\dagger_i = \sum_k c^{(i)}_k \op{P}_k$ as
\begin{widetext}
\begin{equation}
\label{eq:compute-max-fidelity}
\begin{aligned}
	\lvert Z^{0}_{\mathcal{O}}(\vec{\alpha})\rvert 
		&= \lvert \bra{E_n}
			\op{\mathcal{O}}^\dagger(\vec{\alpha})
			\ket{\Omega}\rvert^2
		= \sum_i \left\lvert \alpha_i  \right\rvert^2
		\left\lvert\bra{E_0}\op{\mathcal{O}}_i^\dagger\ket{\Omega}\right\rvert^2 \\
	    &\quad+ 2\mathrm{Re}(\alpha_1 \alpha_2^*)
		\mathrm{Re}\left(
			\bra{E_0}\op{\mathcal{O}}_1^\dagger\ket{\Omega}
			\bra{\Omega}\op{\mathcal{O}}_2\ket{E_0}
		\right) 
	    - 2\mathrm{Im}(\alpha_1 \alpha_2^*)
		\mathrm{Im}\left(
			\bra{E_n}\op{\mathcal{O}}_1^\dagger\ket{\Omega}
			\bra{\Omega}\op{\mathcal{O}}_2\ket{E_n}
		\right) \\
	&= \sum_{k} 
		\left\lvert
			\bra{E_0}\op{P}_k\ket{\Omega}
		\right\rvert^2
		\left(
		\sum_i 
			\lvert c_k^{(i)}\rvert^2
			\lvert \alpha_i  \rvert^2
		\right) \\
	&\quad+ \sum_{k<l}
		\mathrm{Re}
		\left(
			\bra{E_0}
				\op{P}_k
			\ket{\Omega}
			\bra{\Omega}
				\op{P}_l
			\ket{E_0}
		\right)
		\left(
			\sum_i 
				2\mathrm{Re}\left(c^{(i)}_k c^{(i)*}_l\right)
		+ \sum_{i<j} 4\mathrm{Re}
		    \left( 
		        \alpha_i\alpha_j^*c^{(i)}_kc^{(j)*}_l
		    \right)
		\right)\\
		&\quad- \sum_{k<l} 
		\mathrm{Im}
		\left(
			\bra{E_0}
				\op{P}_k
			\ket{\Omega}
			\bra{\Omega}
				\op{P}_l
			\ket{E_0}
		\right)
		\left(
			\sum_i 
				2\mathrm{Im}\left(c^{(i)}_k c^{(i)*}_l\right)
            +\sum_{i<j} 4\mathrm{Im}
                \left(
                    \alpha_i \alpha_j^*c^{(i)}_k c^{(j)*}_l
    			\right)
		    \right).
\end{aligned}
\end{equation}
\end{widetext}
The terms $\lvert \bra{E_0} \op{P}_k\ket{\Omega} \rvert^2$,  $\mathrm{Re}(\bra{E_0} \hat{P}_k \ket{\Omega}\bra{\Omega} \hat{P}_l \ket{E_0})$ and $\mathrm{Im}(\bra{E_0} \hat{P}_k\ket{\Omega}\bra{\Omega} \hat{P}_l \ket{E_0})$ can be estimated with a Hadamard-Overlap test, using controls on the Pauli operators and one ancillary qubit \cite{bravo2019variational}. The overlap test requires preparing both $\ket*{\Omega}$ and $\ket*{E_0}$ states simultaneously. Finally, given all of the measurements, the optimization in Eq.~\eqref{eq:max-fidelity} may be performed classically. Note also that the MF strategy may be generalized to the case when the target state is not a ground state.
\paragraph{Minimum-energy (ME) strategy.} The second strategy is to find $\vec{\alpha}$ that minimizes the energy expectation value of the interpolated state $\ket*{\mathcal{O}(\vec{\alpha})}$ defined in Eq.~\eqref{eq:interpolated-state}, i.e.,
\begin{equation}
\begin{aligned}
\label{eq:min-energy}
\vec{\alpha}_{\mathrm{ME}}
    &= \argmin_{\vec\alpha} 
        \bra{\mathcal{O}(\vec{\alpha})}
            \op{H}
        \ket{\mathcal{O}(\vec{\alpha})} \\
    &= \argmin_{\vec\alpha}
	\frac
		{H_{\mathcal{O}}(\vec\alpha)}
		{N_{\mathcal{O}}(\vec{\alpha})},
\end{aligned}
\end{equation}
where $H_\mathcal{O}(\vec{\alpha}) = \bra{\Omega} \op{\mathcal{O}}(\vec{\alpha})\op{H}\op{\mathcal{O}}^\dagger(\vec{\alpha})\ket{\Omega}$. The effect of this optimization is to minimize excited state contamination in correlation functions at $t=0$ in Euclidean time. 

In implementing the ME strategy, the solution to Eq.~\eqref{eq:min-energy} can be obtained as the eigenvector for the smallest eigenvalue $\lambda_{\mathrm{ME}}$ in a generalized eigenvalue problem (GEVP):
\begin{equation}
\label{eq:me-gevp}
    H_{\mathcal{O}} \vec{\alpha}_{\mathrm{ME}} = \lambda_{\mathrm{ME}} N_{\mathcal{O}}  \vec{\alpha}_{\mathrm{ME}},
\end{equation}
where the matrix pencil is given by the overlap matrix $\lbrack N_{\mathcal{O}} \rbrack_{ij} = \bra{\Omega}\op{\mathcal{O}}_i\op{\mathcal{O}}^\dagger_j \ket{\Omega}$ and the Hamiltonian in the linear subspace spanned by the given interpolating operator set, $\lbrack H_{\mathcal{O}} \rbrack_{ij} = \bra{\Omega}\op{\mathcal{O}}_i\op{H}\op{\mathcal{O}}^\dagger_j \ket{\Omega}$. Here, unlike in classical GEVP optimization of interpolating operators, the matrix elements are local in time. In practice, the vacuum state $\ket*{\Omega}$ may again be approximated with VQE. The matrix elements $\lbrack N_{\mathcal{O}} \rbrack_{ij}$ may be computed using the same measurements as used to compute $N_{\mathcal{O}}(\vec{\alpha})$ in the implementation of the MF strategy. The matrix elements $\lbrack H_{\mathcal{O}} \rbrack_{ij}$ can be computed analogously, by decomposing $\op{\mathcal{O}}_i\op{H}\op{\mathcal{O}}^\dagger_j$ into a basis of Pauli-string operators, and then performing the corresponding Pauli measurements. Once the matrices in the pencil are constructed from measurements on a quantum device, the eigenvector $\vec{\alpha}_{\mathrm{ME}}$ can be obtained in classical post-processing. 

This implementation of the ME strategy is identical to that of quantum subspace expansion (QSE) for electron structure calculations in chemistry applications \cite{PhysRevA.95.042308}. However, unlike in QSE, what is sought here is the eigenvector $\vec{\alpha}_{\mathrm{ME}}$ rather than the eigenstate $\ket*{\mathcal{O}(\vec{\alpha}_{\mathrm{ME}})}$: as discussed in Section~\ref{sec:spectroscopy}, in LQFT the operator construction specified by $\vec{\alpha}_{\mathrm{ME}}$ can be used to find a precise estimate for $E_0$ through a classical LQFT calculation, potentially in calculations in larger lattice volumes and with finer lattice spacings, improving significantly on the upper bound given by the energy of $\ket*{\mathcal{O}(\vec{\alpha}_{\mathrm{ME}})}$ that can be computed within the quantum framework alone.

Compared to maximizing fidelity, minimizing the energy expectation value may yield an operator with a larger overlap onto low-lying excited states which correspond to slowly-decaying excitations in the Euclidean-time correlation function. 
This is a conceptual drawback of the ME strategy which could be a practical concern for systems with closely-spaced low-lying spectra, such as in the case of multi-hadron states in QCD.
However, the implementation of the ME strategy does not require estimating transition probabilities as in Eq.~\eqref{eq:compute-max-fidelity}; only expectation values in the vacuum state are required. This leads to more economical quantum circuits, and the ME strategy is thus more robust against the effects of noise than the MF approach. Noise in quantum circuits results in less-optimal operators and thus additional excited-state contributions in the corresponding correlation functions computed in the Euclidean LQFT. Thus a more robust, though less optimal, strategy may be more practical for applications in the NISQ era.

\section{
    Proof-of-concept implementation for the Schwinger model 
    \label{sec:schwinger}
}
\subsection{
    Schwinger model
    \label{sec:schwinger-background}
}
\par As a proof of concept, both approaches to quantum interpolating operator optimization are implemented for a ground state in the spectrum of the Schwinger model, i.e., quantum electrodynamics in 1+1 dimensions, describing a single species of 2-component spinors $\psi$ with bare mass $m$ coupled to a $U(1)$ gauge field $A_{\mu} = (\phi,A)$ with a dimensionful bare coupling $-g$. In continuum Minkowski spacetime with the metric given by $\eta^{\mu\nu}=\mathrm{diag}(+1,-1)$, 
the corresponding action is given by
\begin{equation}
\begin{aligned}
\label{eq:action}
S &= \int \mathrm{d}^2x\,\left(
	\bar{\psi}
	\left\lbrack 
		i\gamma^\mu D_\mu 
	\right\rbrack
	\psi
	- m \bar{\psi} \psi
	- \frac{1}{4} F_{\mu\nu}F^{\mu\nu}\right)
	,
	%+ \frac{\theta}{2\pi} \oint \mathrm{d}^{\mu}x\,A_{\mu},
\end{aligned}
\end{equation}
where the electric field tensor and the covariant derivative are defined as
\begin{align}
F_{\mu\nu} 
	&= \partial_{\mu} A_{\nu} - \partial_{\nu} A_{\mu}, \label{eq:F-munu}\\
    D_\mu &= \partial_\mu - ig A_{\mu}. \label{eq:covariant-D}
\end{align}
The Clifford algebra of gamma matrices is defined by 
\begin{equation}
\begin{aligned}
\lbrace \gamma^\mu, \gamma^\nu \rbrace &= 2 \eta^{\mu\nu},
 \end{aligned}
 \end{equation}
and chiral projections in two-dimensional Minkowski spacetime require $\gamma_{5} = \gamma_{0} \gamma_1$.
All bound states in the theory are $CP$-even, with quantum numbers $J^{PC}$ either $0^{++}$ or $0^{--}$ \cite{Coleman:1976uz}.
The vacuum state $\ket*{\Omega}$ is the ground state in the even-parity channel. 
The target state for interpolating operator optimization is chosen to be the ground state $\ket*{E_0}$ in the odd-parity channel: the lightest vector meson.

To realize interpolating operator optimization for $\ket*{E_0}$ on a quantum computer, a formulation of the Schwinger model derived from the Kogut-Susskind Hamiltonian is used, following Ref.~\cite{Klco:2018kyo}. 
After a Jordan-Wigner transformation mapping fermion creation and annihilation operators to spin-$1/2$ raising and lowering operators, the (rescaled) Hamiltonian is given by
\begin{equation}
\label{eq:staggered-H}
    \hat{H} = 
     \sum_{n} x
        	\left(\hat{\sigma}^+_n\hat{U}_{1}(n)\hat{\sigma}^-_{n+1} + \mathrm{h.c.}\right)
        + \mu (-1)^n\hat{\sigma}^z_n
        + \hat{\mathcal{E}}^2_n
\end{equation}
(note that the corresponding energy eigenvalues need to be rescaled to express them in lattice units). Here, $\hat{\sigma}^\pm_n$ denote spin raising and lowering operators, which act on site $n$ of the staggered lattice, and $\hat{U}_{1}(n)$ and $\hat{\mathcal{E}}_n$ denote link operators and electric field operators acting on the link connecting staggered sites $n$ and $n+1$. Conventionally, the Hamiltonian is parameterized by two dimensionless quantities, $x= 1/(ag)^2$ and $\mu=2(m_0)/ag^2$, where $a$ is the lattice spacing.

The staggered Kogut-Susskind transformation in 1+1 dimensions with only a spatial discretization completely removes fermion doublers from the spectrum \cite{Kogut:1974ag}. Unlike in higher-dimensional theories \cite{Kluberg-Stern:1983lmr}, the Schwinger model in this formulation contains no additional ``tastes'' of fermions and no taste-breaking effects at finite lattice spacing. In particular, there is a unique vector meson state. Moreover, the spatial boundary conditions in the Kogut-Susskind Hamiltonian are chosen to be periodic (and remain periodic after the Jordan-Wigner transformation if the number of staggered sites is divisible by 4). This allows exact invariance under spatial translations to be preserved, which means that the spatial component of momentum remains a good quantum number. Consequently, to study the vector meson in the rest frame, states with non-zero momentum can be projected out exactly. 

In this prescription, basis elements for the Hilbert space of the Hamiltonian in Eq.~\eqref{eq:staggered-H} are spin chains with $N$ sites and $N$ links, with eigenstates $m_n$ of $\hat{\sigma}^{z}_n$ residing on sites and eigenstates $\ell_n$ of $\hat{\mathcal{E}}_n$ on links, expressed as 
\begin{equation}
\begin{aligned}
\label{eq:basis}
\ket{\Psi}
    &=\ket{\boldsymbol{m}}\otimes\ket{\boldsymbol{\ell}} \\
	&= \ket{m_1, m_{2}, \ldots, m_{N}}\otimes\ket{\ell_1, \ell_{2}, \ldots, \ell_{N}}.
\end{aligned}
\end{equation}
Gauge fixing, necessary in the Hamiltonian formulation, further selects a subspace of physical states. The Kogut-Susskind Hamiltonian is defined in a temporal gauge, $A_0 = 0$; the residual gauge freedom in $\ket{\Psi}$ is fixed by Gauss's law constraints via local symmetry generators $\hat{G}_n$,
\begin{equation}
\label{eq:gauge-constraint}
\hat{G}_{n}\ket{\Psi}=\left(\hat{\mathcal{E}}_{n}-\hat{\mathcal{E}}_{n-1} -\frac{1}{2}\left(\op{\sigma}_n^z + (-1)^n\right)\right)\ket{\Psi} = 0
\end{equation}
at each staggered site $n$, such that after gauge fixing all states contain integer values on each link. Unlike in the gauge-link basis conventionally chosen for classical LQFT calculations, link operators $\hat{U}^\dagger_{1}(n)$ and $\hat{U}_{1}(n)$ in this basis act as raising and lowering operators for the flux quanta $\ell_n$, respectively. 

In this formulation, an interpolating operator optimization problem can be defined as optimizing the linear combination of two zero-momentum, odd-parity interpolating operators for the vector meson:
\begin{align}
\label{eq:interpolator-set-1}
\hat{\mathcal{O}}_{1} 
	&= \sum_{n} 
		(-)^{n}\hat{\mathcal{E}}_n\hat{\mathcal{E}}_{n+2},\\
\label{eq:interpolator-set-2}
\hat{\mathcal{O}}_{2} &= \frac{i}{2}\sum_{n\mathrm{even}} (-)^{n}
		\left(
			\hat{\sigma}^{+}_n\hat{U}_1\hat{\sigma}^{-}_{n+1}
			+ \mathrm{h.c.}
		\right),
\end{align}
where the couplings between every other link in the first operator, and the alternating sign in both operators, are due to the Kogut-Susskind staggered transformation. 
The output of interpolating operator optimization over this set is an eigenvector $\vec{\alpha} = (\alpha_1,\alpha_2)$ specifying the optimal linear combination of operators. The normalization of $\vec{\alpha}$ is irrelevant, so each combination is characterized by the mixing angle 
\begin{equation}
\label{eq:mixing-angle}
    \theta_{\alpha} = \arccos(\alpha_{1}/\lvert      \vec{\alpha} \rvert).
\end{equation}

Of course, many other interpolating operators can be constructed with these quantum numbers and could be included in the optimization set. An artifact of the simple spectrum of the Schwinger model is that it is possible to define simple operator constructions with near-perfect overlap onto the target state; including these in the set would naturally lead to trivial solutions for the vector $\vec{\alpha}$. For numerical demonstration, the interpolating operators are thus chosen so that the overlap factors $\lvert Z^{0}_{\mathcal{O}_1}\rvert$ and $\lvert Z^{0}_{\mathcal{O}_2}\rvert$ are comparable. The parameters $x=4.8$ and $\mu = 0.1$, which correspond to $m_{0}/g \approx 0.18$, are also set specifically to allow a demonstration of non-trivial operator optimization.
\subsection{
    Qubit encoding  
    \label{sec:schwinger-encoding}
}
Following Ref.~\cite{Klco:2018kyo}, the theory's Hilbert space is encoded onto qubits in a quantum simulator. For a discretization with $N=4$ staggered sites, gauge-fixed states are truncated on the allowed values of electric flux such that
\begin{equation}
\begin{aligned}
	\ell_n^2 &\leq \Lambda^2,\qquad
	\sum_{n=1}^{N} \ell_n^2 =\tilde{\Lambda}^2.
\end{aligned}
\end{equation}
The first constraint limits the number of flux quanta on each gauge link, while the second limits their total number in a given state. Retaining only those elements invariant under spatial translations projects onto the zero-momentum subspace. By forming parity eigenstates from the remaining basis elements, projections onto even- and odd-parity sectors are made. These projections block-diagonalize the Hamiltonian in the truncated, zero-momentum subspace of gauge-fixed states; choosing $\Lambda^2=1$ and $\tilde{\Lambda}^2=3$, the basis elements $\ket{\Psi_n}$ in the even parity sector are encoded onto 2-qubit states $\ket{q_{1}q_{2}}$ as
\begin{widetext}
\begin{equation}
\label{eq:even-parity-basis}
\begin{aligned}
\ket{00} &\leftrightarrow \ket{\Psi_1}_{\mathbf{k}=\mathbf{0},\pi+} 
	= \ket{\downarrow \uparrow \downarrow \uparrow}\otimes\ket{0000}, \\
\ket{01} 
	&\leftrightarrow \ket{\Psi_2}_{\mathbf{k}=\mathbf{0},\pi+} 
	=  \frac{1}{2}\left(
		\ket{\uparrow \downarrow \downarrow \uparrow}\otimes\ket{1000}
		+ \ket{\downarrow \downarrow \uparrow \uparrow}\otimes\ket{0\text{-}100}
		+ \ket{\downarrow \uparrow \uparrow \downarrow}\otimes\ket{0010}
		+ \ket{\uparrow \uparrow \downarrow \downarrow}\otimes\ket{000\text{-}1}
        \right),
        \\
\ket{10} &\leftrightarrow \ket{\Psi_3}_{\mathbf{k}=\mathbf{0},\pi+} 
	= \frac{1}{\sqrt{2}}
	\left(
		\ket{\uparrow \downarrow \uparrow \downarrow}\otimes\ket{1010}
		+ \ket{\uparrow \downarrow \uparrow \downarrow}\otimes\ket{0\text{-}10\text{-}1}
     \right),\\
\ket{11} &\leftrightarrow \ket{\Psi_4}_{\mathbf{k}=\mathbf{0},\pi+} 
	= \frac{1}{2}\left(
		\ket{\uparrow \downarrow \downarrow \uparrow}\otimes\ket{0\text{-}1\text{-}1\text{-}1}
		+ \ket{\downarrow \downarrow \uparrow \uparrow}\otimes\ket{1011}
		+ \ket{\downarrow \uparrow \uparrow \downarrow}\otimes\ket{\text{-}1\text{-}10\text{-}1}
		+ \ket{\uparrow \uparrow \downarrow \downarrow}\otimes\ket{1110}
        \right), 
\end{aligned}
\end{equation}
while those in the odd-parity sector are similarly encoded as
\begin{equation}
\label{eq:odd-parity-basis}
\begin{aligned}
\ket{00} &\leftrightarrow \ket{\Psi_1}_{\mathbf{k}=\mathbf{0},\pi-} 
	= \frac{1}{2}\left(
		\ket{\uparrow \downarrow \downarrow \uparrow}\otimes\ket{1000}
		- \ket{\downarrow \downarrow \uparrow \uparrow}\otimes\ket{0\text{-}100}
		+ \ket{\downarrow \uparrow \uparrow \downarrow}\otimes\ket{0010}
		- \ket{\uparrow \uparrow \downarrow \downarrow}\otimes\ket{000\text{-}1}
        \right),
        \\
\ket{01} &\leftrightarrow \ket{\Psi_2}_{\mathbf{k}=\mathbf{0},\pi-} 
	= \frac{1}{\sqrt{2}}
	\left(
		\ket{\uparrow \downarrow \uparrow \downarrow}\otimes\ket{1010}
		- \ket{\uparrow \downarrow \uparrow \downarrow}\otimes\ket{0\text{-}10\text{-}1}
     \right), \\
\ket{10} &\leftrightarrow \ket{\Psi_3}_{\mathbf{k}=\mathbf{0},\pi-} 
	= \frac{1}{2}\left(
		\ket{\uparrow \downarrow \downarrow \uparrow}\otimes\ket{0\text{-}1\text{-}1\text{-}1}
		- \ket{\downarrow \downarrow \uparrow \uparrow}\otimes\ket{1011}
		+ \ket{\downarrow \uparrow \uparrow \downarrow}\otimes\ket{\text{-}1\text{-}10\text{-}1}
		- \ket{\uparrow \uparrow \downarrow \downarrow}\otimes\ket{1110}
        \right)
\end{aligned}
\end{equation}
\end{widetext}
(and the state $\ket{q_1 q_2}=\ket{11}$ is unused). The even and odd-parity blocks in the Hamiltonian are thus given by $4 \times 4$ matrices $\lbrack H_{\mathbf{k=0},\pi\pm}\rbrack_{mn}=  {}_{\mathbf{k=0},\pi\pm}\bra{\Psi_{m}}\hat{H} \ket{\Psi_{n}}{}_{{\mathbf{k=0},\pi\pm}}$, where $\hat{H}$ is the Hamiltonian in Eq.~\eqref{eq:staggered-H}. These matrices are decomposed in a basis of $2$-qubit Pauli-string operators to implement the measurements of expectation values of the Hamiltonian.
\par The interpolating operators in Eqs.~\eqref{eq:interpolator-set-1}--\eqref{eq:interpolator-set-2}, when acting on the vacuum, map from the even-parity sector to the odd-parity sector. Therefore, matrix elements involving interpolated states, such as those defining overlap factors, generally require both sectors of the Hilbert space to be encoded. This is realized by including a third, parity-indicating qubit in the encoding, effectively concatenating the two bases in Eqs.~\eqref{eq:even-parity-basis}--\eqref{eq:odd-parity-basis} as $\ket{\Psi}_{\mom{k=0}}=\ket{\Psi}_{\mom{k=0},\pi+}\otimes\ket{\Psi}_{\mom{k=0},\pi-}$. 
Both strategies for interpolating operator optimization require the $8\times 8$ matrices with matrix elements $\lbrack \mathcal{O}_i\mathcal{O}_j\rbrack_{mn}=  {}_{\mathbf{k=0}}\bra{\Psi_{m}}\op{\mathcal{O}}_i\op{\mathcal{O}}_j \ket{\Psi_{n}}{}_{{\mathbf{k=0}}}$ to be computed; additionally, the MF strategy requires computing
$\lbrack \mathcal{O}_i\rbrack_{mn}=  {}_{\mathbf{k=0}}\bra{\Psi_{m}}\op{\mathcal{O}}_i \ket{\Psi_{n}}{}_{{\mathbf{k=0}}}$, and the ME strategy
$\lbrack \mathcal{O}_i H \mathcal{O}_j\rbrack_{mn}=  {}_{\mathbf{k=0}}\bra{\Psi_{m}}\op{\mathcal{O}}_i\op{H}\op{\mathcal{O}}_j \ket{\Psi_{n}}{}_{{\mathbf{k=0}}}$. The matrices are decomposed in the basis of $3$-qubit Pauli-string operators to implement the requisite measurements, as detailed in Section~\ref{sec:strategies}.
\subsection{
    Numerical demonstration
    \label{sec:schwinger-strategies}
}
\begin{figure*}[!t]
\begin{adjustbox}{width=0.70\textwidth}
\begin{quantikz}
\lstick{} 
    & \gate[wires=3, nwires={2}][1.75cm]{U(\vec{\beta})} 
    & \qw 
    && % equal sign
    & \gate{\exp(-i\beta_2 \frac{\sigma_y}{2})} 
    & \ctrl{2}
    & \qw
    & \ctrl{2}
    & \qw 
    & \rstick{}\qw \\[-0.4cm]
    \lstick{} 
    & % U(beta)
    & % wire
    &=&
    & % gate
    & % control
    & % gate
    & % control
    & % gate
    & \rstick{} \\[-0.4cm]
\lstick{} 
    & % U(beta)
    & \qw
    && % = 
    & \gate{\exp(-i\beta_1 \frac{\sigma_y}{2})}
    & \targ{} 
    & \gate{\exp(-i\beta_1 \frac{\sigma_y}{2})}
    & \targ{}
    & \gate{\exp(-i\beta_3 \frac{\sigma_y}{2})} 
    & \rstick{}\qw \\
\end{quantikz}
\end{adjustbox}
\caption{\label{fig:variational-circuit} A schematic representation of the variational layer in a quantum circuit used to approximate the vacuum $\ket*{\Omega}$ and vector meson $\ket*{E_0}$ states in a VQE for the Schwinger model~\cite{Klco:2018kyo}. The chosen quantum gate basis consists of single-qubit unitaries, represented diagrammatically as boxes of complex matrix exponentials, and two-qubit \textsf{CNOT} gates.
Using this basis, the circuit above efficiently encodes a single-layer variational ansatz $U(\vec{\beta})$, $\vec{\beta}=(\beta_1, \beta_2, \beta_3)$: a general two-qubit unitary transformation subject to the nearest-neighbor structure of the Schwinger-model Hamiltonian in even and odd-parity subspaces of zero momentum~\cite{PhysRevLett.91.147902}.
}
\end{figure*}
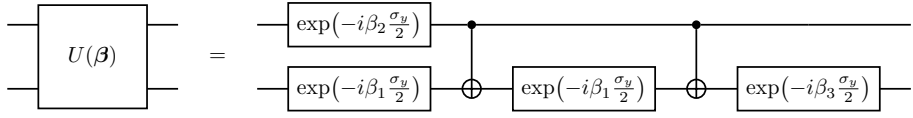
\begin{figure*}[!t]
\begin{adjustbox}{width=0.3\textwidth}
\begin{quantikz}[row sep={0.8cm,between origins}]
\lstick{$q_1 \ket{0}$} 
    & \gate[wires=2][1.75cm]{U(\vec{\beta}^{\pi+})} 
    & \gate[wires=3]{P_k} 
    & \meter{} \\
\lstick{$q_2 \ket{0}$} 
    & % Omega
    & % Pauli_k
    & \meter{} \\
\lstick{$q_3 \ket{0}$} 
    & \qw 
    & % Pauli_k
    & \meter{} 
\end{quantikz}
\end{adjustbox}
\caption{\label{fig:min-energy-schwinger} Diagrammatic representation of quantum circuits used to estimate $H_\mathcal{O}(\vec{\alpha}) = \bra{\Omega} \op{\mathcal{O}}(\vec{\alpha})\op{H}\op{\mathcal{O}}^\dagger(\vec{\alpha})\ket{\Omega}$ for the minimum-energy strategy and $N_{\mathcal{O}}(\vec{\alpha}) =  \bra{\Omega}\op{\mathcal{O}}(\vec{\alpha})\op{\mathcal{O}}^\dagger(\vec{\alpha})\ket{\Omega}$ for both strategies by measuring $\bra*{\widetilde{\Omega}} \hat{P}_k \ket*{\widetilde{\Omega}}$. Here, $\hat{P}_k$ are 3-qubit Pauli-string operators entering the decomposition of $\lbrack \mathcal{O}_i \mathcal{O}_j \rbrack$ or $\lbrack \mathcal{O}_i H \mathcal{O}_j \rbrack$, with $i,j\in\lbrace 1,2\rbrace$. First, a quantum state $\ket*{\widetilde{\Omega}}$ is prepared from the fiducical state $\ket*{q_1 q_2 q_3} = \ket*{000}$ by acting with a variational ansatz $U(\bm{\beta}^{\pi+})$ on $\ket{q_1 q_2}$, followed by applying $\hat{P}_k$ on $\ket{q_1 q_2 q_3}$; followed by measurement gates.
}
\end{figure*}
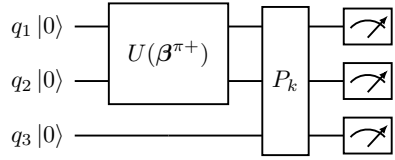
\begin{figure*}[!t]
\begin{adjustbox}{width=0.5\textwidth}
\begin{quantikz}[row sep={0.8cm,between origins}]
\lstick{$a_1 \ket{0}$} 
    & \gate{H}
    & \ctrl{1} % Pauli on q_{1}
    & \ctrl{4} % Pauli on q_{n+1}
    & \gate[style={fill=blue!20}]{R_z} % control 1
    & \qw % control 2
    & \qw % control 3
    & \gate{H}
    & \meter{} \\
\lstick{$q_1 \ket{0}$} 
    & \gate[wires=2][1.75cm]{U(\vec{\beta}^{\pi+})} 
    & \gate[wires=3]{P_k} 
    & \qw % P_l 
    & \ctrl{3}
    & \qw % control 2
    & \qw % control 3
    & \gate{H} 
    & \meter{} \\
\lstick{$q_2 \ket{0}$} 
    & % Omega
    & % Pauli_k
    & \qw % Pauli_l
    & \qw % control 1
    & \ctrl{3} % control 2
    & \qw % control 3
    & \gate{H} 
    & \meter{} \\
\lstick{$q_3 \ket{0}$} 
    & \qw 
    & % Pauli_k
    & \qw % Pauli_l
    & \qw % control 1
    & \qw % control 2
    & \ctrl{3} % control 3
    & \gate{H} 
    & \meter{} \\
\lstick{$q_{4} \ket{0}$} 
    & \gate[wires=2][1.75cm]{U(\vec{\beta}^{\pi-})} 
    & \qw % Pauli_l
    & \gate[wires=3]{P_l} 
    & \targ{} \qw
    & \qw % control 2
    & \qw % control 3
    & \gate{H} 
    & \meter{} \\
\lstick{$q_{5} \ket{0}$} 
    & % Omega
    & \qw % Pauli_k
    & % Pauli_{l}
    & \qw % control 1
    & \targ{}
    & \qw % control 3
    & \gate{H} & \meter{} \\
\lstick{$q_6 \ket{0}$} 
    & \gate{X}
    & \qw
    & % Pauli_l
    & \qw % control 1
    & \qw % control 2
    & \targ{} \qw 
    & \gate{H} 
    & \meter{} 
\end{quantikz}
\end{adjustbox}
\caption{\label{fig:hadamard-overlap-schwinger} Diagrammatic representation of quantum circuits for the Hadamard-overlap test \cite{bravo2019variational} used to estimate the overlap factor $\lvert Z^0_{\mathcal{O}}(\vec{\alpha}) \rvert = \lvert \bra{E_0}\op{\mathcal{O}}^\dagger(\vec{\alpha})\ket{\Omega} \rvert^2$ for the maximum-fidelity strategy. 
The circuits measure $\lvert \bra*{\widetilde{E}_0}\op{P}_k\ket*{\widetilde{\Omega}}\rvert^2$,  $\mathrm{Re}(\bra*{\widetilde{E}_0}\hat{P}_k\ket*{\widetilde{\Omega}}\bra*{\widetilde{\Omega}}\hat{P}_l\ket*{\widetilde{E}_0})$ and $\mathrm{Im}(\bra*{\widetilde{E}_0}\hat{P}_k\ket*{\widetilde{\Omega}}\bra*{\widetilde{\Omega}}\hat{P}_l\ket*{\widetilde{E}_0})$ entering  Eq.~\eqref{eq:compute-max-fidelity}. 
First, quantum states $\ket*{\widetilde{\Omega}}$ and $\ket*{\widetilde{E}_0}$ are prepared with variational ans{\"a}tze $U(\bm{\beta}^{\pi\pm})$ and the single-qubit Pauli gate $X$, and the ancillary qubit $a_1$ is put in an equal superposition of $0$ and $1$ with the Hadamard gate $H$. 
Then, three-qubit Pauli gates are applied conditional on $a_1=1$. Finally, a $\textsf{SWAP}$ test is performed using \textsf{CNOT}, Hadamard, and measurement gates.
The colored $R_{z}$ gate on the ancillary qubit denotes the rotation about the $\hat{z}$-axis by an angle of $-\pi/2$. For $k \neq l$, measurements for the real component of  $\bra*{\widetilde{E}_0}\hat{P}_k\ket*{\widetilde{\Omega}}\bra*{\widetilde{\Omega}}\hat{P}_l\ket*{\widetilde{E}_0})$ are made when the $R_{z}$ gate is omitted; for the imaginary component, when it is applied.}
\end{figure*}
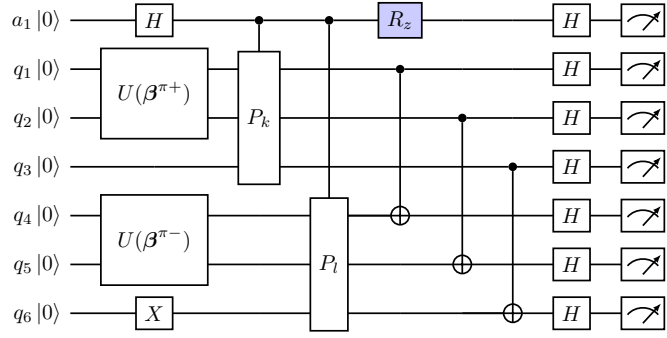
\par A quantum state approximating the even-parity ground state $\ket*{\Omega}$ is required for both ME and MF interpolating operator optimization strategies, while the MF strategy also requires a quantum state to approximate the odd-parity ground state $\ket*{E_0}$. These states are prepared using VQE on 2-qubit variational circuits operating within the even and odd-parity sectors encoded as described in Eqs.~\eqref{eq:even-parity-basis} and \eqref{eq:odd-parity-basis}, respectively, to yield approximate eigenstates $\ket*{\widetilde{\Omega}}$ and $\ket*{\widetilde{E}_0}$. A single variational layer is used for both circuits, so that each state preparation is specified by a vector of variational parameters $\vec{\beta}^{\pi\pm} = (\beta^{\pi\pm}_1,\beta^{\pi\pm}_2,\beta^{\pi\pm}_3)$, as illustrated in Fig.~\ref{fig:variational-circuit}. To iteratively search for $\vec{\beta}^{\pi\pm}$, the first-order simultaneous perturbation stochastic approximation (SPSA) optimizer is used \cite{Spall1998ANOO}, with default settings within the \textsf{Qiskit} software development kit \cite{Qiskit} (metapackage version \textsf{0.29.0} \cite{qiskitversions, qiskitSPSAversion}).
Once the optimal variational parameters are found, the corresponding states $\ket*{\widetilde{\Omega}}$ and $\ket*{\widetilde{E}_0}$ are re-prepared to perform the requisite matrix element measurements for both strategies. 
\begin{figure*}[!t]
\centering
\includegraphics[scale=0.7]{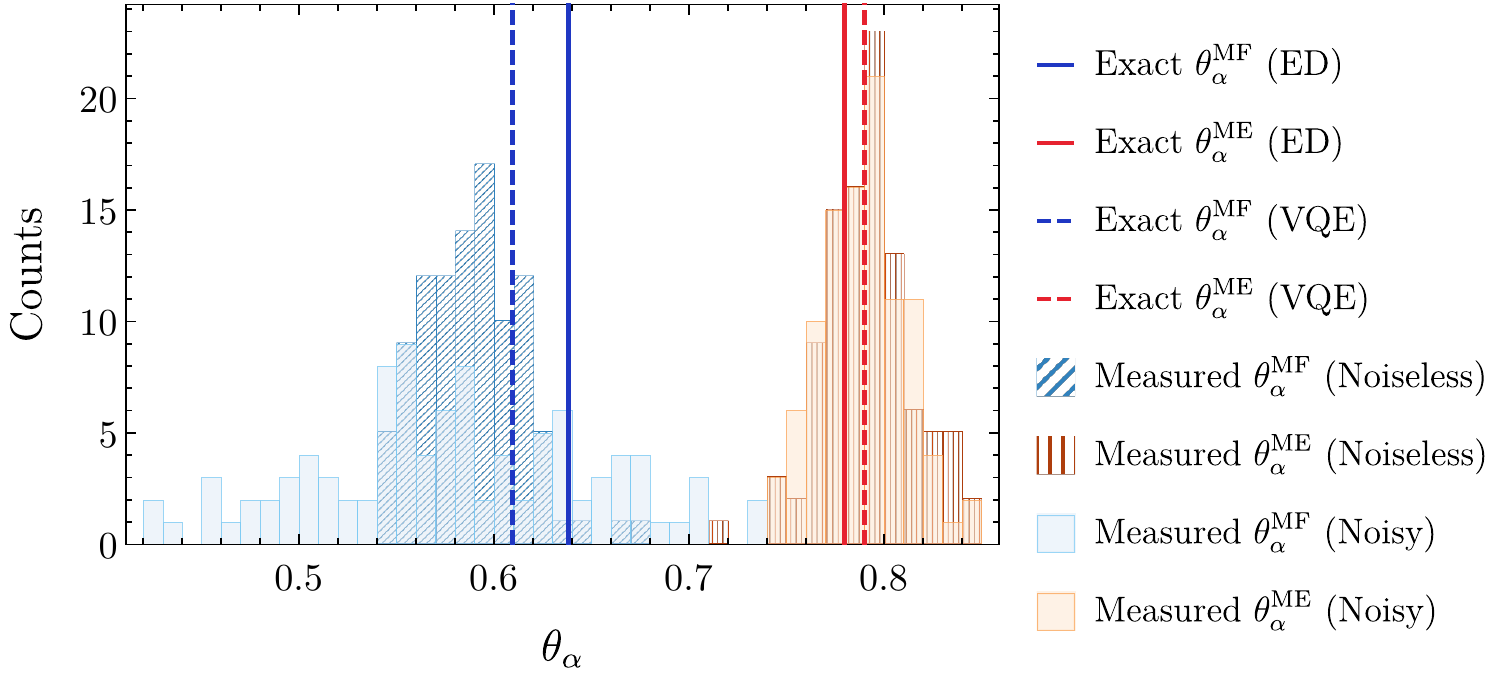}
\caption{\label{fig:mixing-angles}
Histogram of the optimized mixing angles $\theta_\alpha$ determined using the minimum-energy (orange) and the maximum-fidelity (blue) strategies. 
The distribution in the mixing angle induced by finite statistics (hatched), and both finite statistics and the effects of noise (solid) are shown. 
All histograms show 100 values of the mixing angle obtained from an ensemble of sets of 8192-shot measurements. 
The means of the distributions are biased, as illustrated by the difference between the solid and dashed vertical lines, because the variational approximations $\ket*{\widetilde{\Omega}}$ and $\ket*{\widetilde{E}_0)}$ do not exactly reproduce the exact ground states $\ket*{\Omega}$ and $\ket*{E_0}$. 
The positions of the solid lines are computed by exact diagonalization (ED), the dashed lines using the \textsf{StateVector} simulator in the \textsf{Qiskit} software development kit, which removes the effects of finite statistics.}
\end{figure*}

\par For the maximum-fidelity strategy, $\theta_{\alpha}^{\mathrm{MF}}$ is obtained from Eqs.~\eqref{eq:max-fidelity} and \eqref{eq:mixing-angle} given measurements of $\bra*{\widetilde{\Omega}} \hat{P}_k \ket*{\widetilde{\Omega}}$ for the interpolated state normalization $N_{\mathcal{O}}(\vec{\alpha})$ in the denominator, and $\lvert \bra*{\widetilde{E}_0}\op{P}_k\ket*{\widetilde{\Omega}} \rvert^2$,  $\mathrm{Re}(\bra*{\widetilde{E}_0}\hat{P}_k\ket*{\widetilde{\Omega}}\bra*{\widetilde{\Omega}}\hat{P}_l\ket*{\widetilde{E}_0})$, and $\mathrm{Im}(\bra*{\widetilde{E}_0}\hat{P}_k\ket*{\widetilde{\Omega}}\bra*{\widetilde{\Omega}}\hat{P}_l\ket*{\widetilde{E}_0})$ for the overlap factor $\lvert Z^0_{\mathcal{O}}(\vec{\alpha})\rvert$ in the numerator. 
Measurements of diagonal elements $\bra{\Omega} \hat{P}_k \ket{\Omega}$ are performed with 3-qubit circuits as shown in Fig.~\ref{fig:min-energy-schwinger}; the remaining off-diagonal elements are measured with $7$-qubit circuits as shown in Fig.~\ref{fig:hadamard-overlap-schwinger}, following the Hadamard-overlap test.
All matrix elements are computed using an ensemble of 100 sets of measurements, each measurement with 8192 shots, using the quantum assembly language (\textsf{QASM}) \cite{Cross:2021bcz} simulator provided through IBM's Quantum Experience \cite{IBMQuantum} and the \textsf{Qiskit} software development kit \cite{Qiskit}. 
To simulate the effect of gate errors, the simulator is provided with basis gates, the qubit coupling map, and the noise model from a real quantum device; in this work, the 7-qubit \textsf{ibmq\_casablanca} device is used, which was one of the IBM Quantum Falcon processors (now retired).

\par For the Minimum-Energy strategy, $\theta_{\alpha}^{\mathrm{ME}}$ is obtained from Eqs.~\eqref{eq:min-energy} and \eqref{eq:mixing-angle}. As described in Section~\ref{sec:strategies},
the corresponding eigenvector $\vec{\alpha}$ is a solution to a generalized eigenvalue problem with the pencil of $2\times 2$ matrices  $\lbrack N_{\mathcal{O}} \rbrack_{ij} = \bra*{\Omega}\op{\mathcal{O}}_i\op{\mathcal{O}}^\dagger_j \ket*{\Omega}$ and $\lbrack H_{\mathcal{O}} \rbrack_{ij} = \bra*{\Omega}\op{\mathcal{O}}_i\op{H}\op{\mathcal{O}}^\dagger_j \ket*{\Omega}$. The matrix elements are estimated in the same way as those for the MF strategy. 

After optimization, each set of measurements yields a mixing angle describing an optimized combination of interpolating operators for a each strategy, and the ensemble of measurements samples from the corresponding distribution of mixing angles. These distributions, as well as exact values of the mixing angles from the eigenstates and their variational approximations, are shown for both strategies in Fig.~\ref{fig:mixing-angles}. Clearly, even in this simple implementation, the two strategies lead to different optimized interpolating operator constructions.

Zero-momentum two-point correlation functions $C(\mom{0},t)$ are computed from each optimized interpolating operator, as well as from the two unoptimized constructions as a benchmark. For this proof-of-principle demonstration, only the eigenvectors $\vec{\alpha}$ come from quantum simulation data; the remaining components --- eigenenergies $E_n$ and overlap factors $\lvert Z^{n}_{\mathcal{O}}(\vec{\alpha})\rvert$ given $\vec{\alpha}$ --- are found by exact diagonalization, as a surrogate for the expectation value of their stochastic estimates computed in classical LQFT calculations. 

\begin{figure*}[!t]
\centering
\includegraphics[scale=0.7]{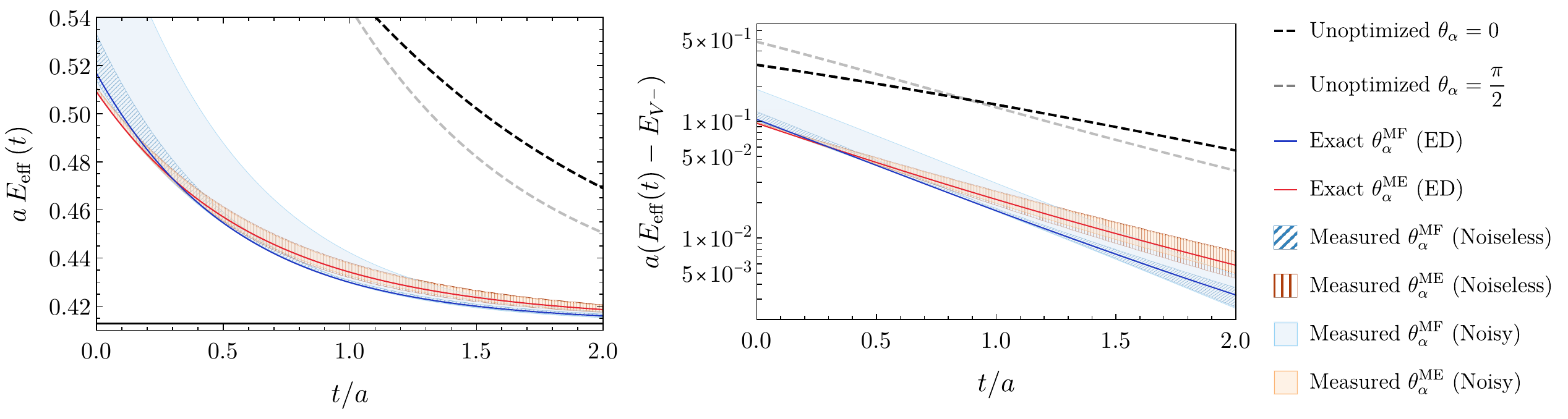}
\caption{
Comparison of effective energy functions (Eq.~\eqref{eq:effective-energy}) computed from correlation functions constructed with unoptimized and optimized interpolating operators, with and without the effects of noise in the quantum device. 
Unoptimized constructions use the operators defined in Eq.~\eqref{eq:interpolator-set-1} (black) and Eq.~\eqref{eq:interpolator-set-2} (gray). Curves labeled ``Exact'' correspond to optimized constructions with mixing angles obtained from exact diagonalization rather than quantum simulation measurements. For each quantum optimization strategy, hatched (solid) bands labeled ``Measured'' encompass 100 curves obtained from multi-shot measurements in quantum simulation without (with) noise. 
The left panel shows the effective energy functions and their asymptote, the mass of the vector meson state from exact diagonalization (ED); the right panel shows the excited state contamination in each function on a logarithmic scale. Energies are converted to lattice units from the units of the rescaled Hamiltonian in Eq.~\eqref{eq:staggered-H}.
\label{fig:mass-curves}}
\end{figure*}

Effective energy functions computed from the correlation functions (Eq.~\eqref{eq:effective-energy}) are shown in Fig.~\ref{fig:mass-curves}. The bands shown span the range of all 100 measurement sets with each strategy. 
Clearly, constructions optimized following both strategies result in substantially less excited state contamination compared to unoptimized constructions (i.e., with $\theta_{\alpha} = 0$ and $\theta_{\alpha} = \pi/2$). 
These improvements persist despite the effects of the VQE approximations, finite statistics, and gate errors.
required eigenstates. 
The differences between the exact results of the two strategies are also as expected; the ME approach results in smaller excited state contamination at zero Euclidean time, but a somewhat longer Euclidean time evolution is required to suppress those contributions. In contrast, while the initial excited state contamination is larger from the MF approach, those contributions are suppressed more rapidly in Euclidean time. Thus, when ignoring the effects of noise on the quantum device, optimizing interpolating operator constructions by maximizing fidelity is preferable for the purposes of lattice spectroscopy.

However, gate errors have a substantially different impact on the two strategies. Since the implementation of the ME approach can be achieved with more economical circuits than the MF approach,
%with more economical circuits than the MF approach. The former strategy requires quantum circuits no more complicated than those corresponding to the VQE ansatz for the vacuum state. 
gate noise has a smaller effect on the width of the corresponding mixing-angle distribution, as illustrated in Fig.~\ref{fig:mixing-angles}. The practical significance of those differences for LQFT calculations can be judged by comparing how effective energy curves from each strategy are affected by noise. Effective energies for the ME-optimized interpolating operator with and without simulated gate noise are practically indistinguishable in the left panel of Fig.~\ref{fig:mass-curves}, with small differences that can only be seen in the figure with a logarithmic scale in the right panel. In contrast, adding the effects noise for MF-optimized interpolating operators significantly increases the width of the effective energy distribution, producing a broader band (whose width is nonetheless suppressed in Euclidean time, as expected).

Other, less significant, differences between the effective energy curves obtained from exact matrix elements and those estimated from measurements in a quantum simulator are also evident for each strategy; although not practically significant for the current case, they reveal differences between the two strategies that may be important in applications to other systems. 
For the ME strategy, the measurement set can sometimes lead to a faster decay of the effective energy than that resulting from the exact solution.
This is possible because the ME solution minimizes the energy expectation value at zero Euclidean time, which does not guarantee the fastest rate of decay.
As a result, the effects of finite statistics and noise can be observed in the widening of ME bands both above and below the exact ME curve at late times. 
In contrast, for the MF strategy, noise may lead to fluctuations below the exact effective energy at $t=0$, where the two MF bands contain regions both above and below the exact MF curve, but their effect on the rate of decay is more skewed toward decreasing the convergence to the target, thus leading to a widening of MF bands primarily above the exact solution at late Euclidean times.

Finally, in practice the optimal choice of strategy also depends on how late in Euclidean time the signal can be extracted.
This effect is not illustrated in this proof-of-principle study, where correlation functions are computed exactly, but will become important when they are estimated from Monte Carlo data and the range of a statistical signal in Euclidean time is limited by the diminishing signal-to-noise ratio. 
Depending on the onset of this stochastic noise, the faster convergence of the MF strategy could be practically significant. In that case, a combination of the two strategies could be used to mitigate the effects of both kinds of noise: noise in the quantum device and stochastic noise from Monte Carlo estimates. 
\section{
    Hybrid quantum-classical workflow
    \label{sec:hybrid-workflow}
}
The discussion of Section~\ref{sec:schwinger-strategies} comparing the two different approaches to quantum optimization of interpolating operators is concerned only with the implementation of the strategies on a quantum device, and not with the translation of optimized constructions to classical LQFT. Of course, within the proposed hybrid quantum-classical workflow, correlation functions from optimized interpolating operator constructions would not be computed from exact diagonalization, but rather would be estimated from field configurations sampled with Monte Carlo methods in a classical LQFT calculation. This is somewhat complicated by the different formulations of the same LQFT which may be used in quantum and classical calculations. Classical LQFT calculations conventionally utilize Lagrangian formulations in position space, in discrete Euclidean time, with one of the several established fermion discretizaton schemes and gauge fields represented in terms of link variables. If one of these formulations is chosen for the classical portion of the calculation, a number of constraints on the formulation used for the quantum calculation need to be considered to realize the proposed hybrid workflow.

First, Hamiltonian formulations in quantum computation may be thought of as a temporal continuum limit of Lagrangian formulations in LQFT \cite{Loan:2002ej, Byrnes:2003gg}. Taking this limit renormalizes the couplings and masses, raising the need to match couplings and parameters in the two formulations of the theory \cite{Kogut:1979vg, Dashen:1980vm, Hasenfratz:1981tw, Karsch:1982ve}. Ignoring the running of couplings could render the interpolating operator constructions optimized via quantum computation less effective, as the constructions would essentially have been optimized and applied for theories at different physical parameters.
 
Moreover, the choice of basis used to represent quantum fields has an effect on what operators can be practically implemented in a calculation. One relevant example is that typical interpolating operators used in classical LQFT calculations are non-unitary; yet, additional challenges are posed in realizing such operators with quantum gates. Furthermore, the choice of basis also affects the variance in the stochastic estimate of a correlation function from  a given interpolating operator. This is similar in spirit to the observation that the choice of interpolating operator set affects the variance \cite{Detmold:2014hla}, and directly related to numerical demonstrations of reducing variance in LQFT observables by path integral contour deformations \cite{Aarts:2008rr, Cristoforetti:2012su, Schmidt:2017gvu, DiRenzo:2017igr, Alexandru:2018fqp, Zambello:2018ibq, Ohnishi:2019ljc, Detmold:2020ncp, Lawrence:2020kyw, Detmold:2021ulb,  Alexandru:2020wrj}. For this reason, an interpolating operator, optimized in a quantum computation in one basis, may suppress excited state contamination in the correlation function in a study within the same formulation, but a stochastic estimate of the corresponding correlation function sampled in LQFT with Monte Carlo with gauge fields represented in a different basis may be rendered impractical by signal-to-noise problems. 

Despite these challenges, matching between two formulations in the proposed hybrid workflow is also a key advantage of the approach when considering the use of NISQ-era quantum technology.
Namely, thanks to the finite spatial extent of bound states, it is to be expected that effective interpolating operator constructions may be obtained from optimization performed for a smaller physical volume than in a target classical calculation. 
A quantum calculation in a smaller volume would require less computational resources, but could still be used to increase the computational efficiency of the principal, classical stage of the hybrid workflow.

This robustness makes interpolating operator optimization particularly suitable for NISQ-era technology.
Moreover, LQFT calculations that require lattice volumes exceeding NISQ-era capabilities, for example those encoding the interactions of the target state with external probes via current insertions, can be also made more efficient on classical computers with interpolating-operator constructions optimized in small-volume quantum calculations.
\section{
    Conclusions and outlook 
    \label{sec:conclusion}
}
In summary, interpolating operator optimization for LQFT calculations is a suitable problem for NISQ-era technology that could be handled as a pre-computation task in a hybrid quantum-classical LQFT workflow, as proposed originally in Ref~\cite{Avkhadiev:2019niu}. This work presents two concrete strategies, and their proof-of-concept realization, for interpolating operator optimization on a quantum computer: the ``MF" strategy of maximizing fidelity of a normalized interpolated state to the target eigenstate, and the ``ME" approach of minimizing the energy expectation value of a normalized interpolated state. The first strategy is generalizable to target states other than ground states, and is preferable in the absence of gate errors as it directly minimizes excited-state contributions in the spectral decomposition of a two-point correlation function. This ensures a quicker decay of excited-state contamination of correlation functions in Euclidean time relative to that from unoptimized constructions. In contrast, the second strategy minimizes excited-state contributions at zero Euclidean time. Compared to the first strategy, this approach may lead to larger overlaps onto excited-state contributions with lower energies, potentially resulting in persistent slowly-decaying modes in the correlation function. In the implementation of these approaches for a vector meson state in the Schwinger model, such behavior is not found to be significant. Moreover, when gate errors based on current real-device noise models are added to the simulator, the ME strategy is shown to be more robust to noise than the MF strategy, whose implementation requires deeper quantum circuits with more entangling gates and a larger number of qubits.

Practical implementations of interpolating operator optimization strategies in the future could be improved by considering alternative encodings of typical LQFT interpolating operators. This could be achieved, for example, by block-encoding the interpolating operators: representing the non-unitary operators as blocks in a larger unitary matrix \cite{gilyen2019quantum}. Alternatively, the interpolation between the vacuum and the target ground state could be first found as unitary quantum circuit, and then approximated by a combination of non-unitary interpolating operators that can be practically realized in LQFT. 

Regardless of a particular strategy or implementation, interpolating operator optimization with quantum-computer technologies is different in spirit from classical optimization approaches in LQFT, in that it deals directly with representations of quantum states rather than stochastic estimates of the corresponding correlation functions. In particular, this means that quantum interpolating operator optimization is not hindered by the exponential suppression of the signal-to-noise ratio in Euclidean time from Monte Carlo data. This feature makes interpolating operator optimization on quantum computers an exciting direction for further research. 

Finally, this work discusses how matching between the formulations of quantum and classical calculations for the same theory may enable interpolating operator optimization with NISQ-era devices in a hybrid quantum-classical workflow.
In this approach, interpolating operator constructions, optimized in auxiliary small-scale quantum calculations designed to study isolated bound states, may increase the precision at fixed statistics, and thus accelerate, principal large-scale LQFT calculations on classical computers. These classical calculations may, for example, study interactions that include the same states and perform continuum extrapolations. Crucially, by virtue of Euclidean time evolution, the principal classical calculations will be robust to noise that enters quantum-computer calculations as well as the systematic uncertainties that arise from matching between the two calculations performed at different physical volumes, lattice spacings, and truncations on the gauge field space (in the quantum calculation). In this way, quantum interpolating operator optimization presents a potential use case for NISQ-era technology for LQFT before standalone LQFT calculations on quantum computers are practical at scale. 
\begin{acknowledgements} 
The authors thank Will Detmold and Lena Funcke for helpful discussions.
This work is supported in part by the U.S.~Department of Energy, Office of Science, Office of Nuclear Physics under grant Contract Number DE-SC0011090 and by the National Quantum Information Science Research Centers, Co-design Center for Quantum Advantage (C\textsuperscript{2}QA) under contract number DE-SC0012704. PES is additionally supported by the National Science Foundation under EAGER grant 2035015, and by the U.S. DOE Early Career Award DE-SC0021006. RDY is supported by the Australian Research Council grants DP190100297 and DP220103098.
\par The authors acknowledge the use of IBM Quantum services for this work. The views expressed are those of the authors, and do not reflect the official policy or position of IBM or the IBM Quantum team.
\end{acknowledgements} 
\bibliography{refs.bib}
\end{document}